# UNDULATOR-TYPE RADIATION OF BUNCHED CHARGED PARTICLES IN SELF-WAKEFIELD [*]

Anatoliy Opanasenko[**]

NSC KIPT, Akademicheskaya St. 1, Kharkov, 61108, Ukraine

Radiation appearing when relativistic charged particles moves along a periodic structure without external fields is investigated. It is shown that nonsynchronous spatial harmonics of wakefields excited by bunched charged particles can give rise to the particle oscillatory motion that consequently generates the undulator-type radiation (UR). A theory of the undulator-type radiation emitted by ultrarelativistic charged particles in the self-wakefields is given. An analytical expression for the spontaneous UR power of the ultrarelativistic monochromatic charged bunch moving in a weakly corrugated axially-symmetrical waveguide is derived by the perturbation method. The parameter region, a particle number and particle energies at which the spontaneous UR power exceeds the wakefield power is analyzed.

**1.** As is known, a charged particle bunch moving with a constant velocity along a periodic structure emits the Cherenkov-type radiation (or the diffraction radiation) [1]. The fields of this radiation called wakefields can be expressed as a spatial harmonic series according to the Floquet's theorem. Action of the synchronous spatial harmonics of the wakefield on the particles results in energy losses associated with the Cherenkov-type radiation. As is well known, the nonsynchronous spatial harmonics do not contribute to an energy change of a beam on the average over the structure period $D$. However under certain conditions, the nonsynchronous spatial harmonics give rise to the oscillatory motion of the particles with a time period equal to the flight time of the structure period. That, consequently, can provide the generation of the undulator-type radiation. This radiation is the subject of discussion in this article.

**2.** At the beginning we will consider the radiation of a self-oscillating point charge in the periodic structure [2]. As a periodic structure, we choose the vacuum corrugated waveguide with a metallic surface. Such structures are commonly used in rf linacs. Let a particle having the ultrarelativistic velocity $\mathbf{v}$, the charge $e$ and the mass $m$ moves along the structure with the period $D$. The longitudinal component of the velocity $v_z$, parallel to the structure axis, is very close to the velocity of light $c$. The radiation reaction force and the radiation power have to be found.

Using the Hamilton's method developed in Ref. [3] we can obtain the radiation reaction force as

$$\mathbf{F}(\mathbf{v}(t),\mathbf{r}(t),t) = -\frac{e^2}{4c^2 V_{tot}} \sum_\lambda^{\omega_\lambda < c/r_0} \left\{ \left[ \mathbf{A}_\lambda(\mathbf{r}(t)) - \frac{\mathbf{v}(t) \times rot \mathbf{A}_\lambda(\mathbf{r}(t))}{i\omega_\lambda} \right] e^{i\omega_\lambda t} \int_0^t \mathbf{v}(t') \mathbf{A}_\lambda^*(\mathbf{r}(t')) e^{-i\omega_\lambda t'} dt' + \left[ \mathbf{A}_\lambda(\mathbf{r}(t)) + \frac{\mathbf{v}(t) \times rot \mathbf{A}_\lambda(\mathbf{r}(t))}{i\omega_\lambda} \right] e^{-i\omega_\lambda t} \int_0^t \mathbf{v}(t') \mathbf{A}_\lambda^*(\mathbf{r}(t')) e^{i\omega_\lambda t'} dt' \right\} + c.c. \quad (1)$$

where $\omega_\lambda$ are the eigenfrequencies. Seeing the radiation reaction force does not depend on the particle size $r_0$, so $\omega_\lambda < c/r_0$ [3]. $V_{tot} = MV_{cell}$, here we assume that the structure contains $M \to \infty$ cells of a volume $V_{cell}$ and is enclosed in a "periodicity box". $\mathbf{A}_\lambda(\mathbf{r})$ is the set of the eigenfunctions of the vector potential which can be represented in the Floquet form [1]

$$\mathbf{A}_\lambda(\mathbf{r}) = \sum_{n=-\infty}^{\infty} \mathbf{g}_\lambda^{(n)}(\mathbf{r}_\perp) e^{ih_n z}, \quad (2)$$

where $\mathbf{g}^{(n)}_\lambda(\mathbf{r}_\perp)$ is the amplitude of the $n$th spatial harmonic; $h_n = h + 2\pi n/D$ is the propagation constant of the $n$th spatial harmonic; $h$ is the discrete parameter multiple of $2\pi/(MD)$ in the interval $(-\pi/D \div \pi/D)$.

The set of eigenfunctions of Eq.(2) for the infinitely long periodic waveguide is physically limited in frequency by the value of the electron plasma frequency $\omega_{pe}$ in the metal. As is known, if $\omega_\lambda \sim \omega_{pe}$, the conduction of metal walls strongly falls off, and the diffraction conditions in the periodic structure are disrupted. So, in the spectral region $\omega_\lambda > \omega_{pe}$, where the wave diffraction can be neglected, the periodic waveguide can be considered as a free space. In this part of the frequency spectrum, the vector potential is sought as an expansion in terms of the plane waves

$$\mathbf{A}_{\lambda,l}(\mathbf{r}) = c\sqrt{4\pi}\,\boldsymbol{a}_{\lambda,l} e^{i\mathbf{k}_\lambda \mathbf{r}}, \quad (3)$$

where $\boldsymbol{k}_\lambda$ is the wave propagation vector; $\boldsymbol{a}_{\lambda,l}$ are the real unit vectors of polarization ($l=1,2$) perpendicular to $\boldsymbol{k}_\lambda$.

In the ultrarelativistic limit, the equation of motion

---

$$\frac{d}{dt}\frac{m\mathbf{v}}{\sqrt{1-v^2/c^2}} = \mathbf{F}(\mathbf{v}(t),\mathbf{r}(t),t) \tag{4}$$

driven by the radiation reaction force can be solved by the method of successive approximations. We will find non-relativistic corrections for the particle velocity $v_0 \approx c$. As a zeroth order approximation, we consider the uniform motion of the charged particle parallel to the waveguide axis

$$\mathbf{v} = \mathbf{v}_0 = v_0 \mathbf{e}_z, \qquad \mathbf{r}(t) = \mathbf{r}_{0,\perp} + \mathbf{v}_0 t. \tag{5}$$

Inserting Eqs. (5) and (2) (for the frequency region $\omega_\lambda << \omega_{pe}$) into Eq. (1) we obtain the self-wake force in the zeroth-order approximation

$$\mathbf{F}(t) = -e^2 \sum_{p=-\infty}^{\infty} \mathbf{u}^{(p)} e^{ip\Omega t}, \tag{6}$$

where $\Omega \equiv 2\pi v_0/D$ and the amplitudes of the $p$th spatial harmonics of the self-wake function are defined as

$$\mathbf{u}^{(p)} = \mathbf{w}^{(p)} + \mathbf{w}^{(-p)*}, \tag{7}$$

$$\mathbf{w}^{(p)} \equiv \frac{Dv_0}{4c^2 V_{cell}} \sum_{n=0}^{\infty} \sum_{\lambda'_j} \frac{g_{z,\lambda_j}^{(n)*}}{\left|v_0 - \frac{d\omega_\lambda}{dh}\right|_{\lambda=\lambda_j}} \left[ \mathbf{g}_{z,\lambda_j'}^{(n+p)} - i\frac{v_0}{\omega_\lambda} \nabla_\perp g_{z,\lambda_j}^{(n+p)} - \frac{\Omega p}{\omega_\lambda} \mathbf{g}_{\perp,\lambda_j}^{(n+p)} \right].$$

Hereinafter, the amplitudes of spatial harmonics are taken at $\mathbf{r}=\mathbf{r}_{0,\perp}$ as $\mathbf{g}_\lambda^{(n)} \equiv \mathbf{g}_\lambda^{(n)}(\mathbf{r}_{0,\perp})$, $\omega_{\lambda_j}$ satisfies the resonance conditions $\omega_\lambda - hv_0 = n\Omega$.

The wake force (6) is the periodic function of time with the period $D/v_0$. The synchronous harmonic of the force $-e^2 2w_z^{(0)}$ defines the energy losses associated with Cherenkov-type radiation in the range $\omega_\lambda << \omega_{pe}$. As it is easily seen, the transverse component of the synchronous harmonic of the self-wake force equals zero $\mathbf{w}_\perp^{(0)}=0$.

If a charged particle moves off-axis, it experiences the action of the transverse component of nonsynchronous harmonics of the self-wake force ($\mathbf{w}_\perp^{(p)} \neq 0$). So, we will find non-relativistic corrections for both the velocity and the radius vector of the off-axis particle that are caused by the periodic transverse self-wake force. We assume that the change in the longitudinal velocity is negligible. Solving the equation of the motion driven by the force Eq.(6) we correct the law of motion

$$\mathbf{v}(t) = \mathbf{v}_0 + \mathbf{v}_\perp(t) = \mathbf{v}_0 + ic\sum_{p\neq 0}^{\infty} \frac{\boldsymbol{\alpha}^{(p)}}{p} e^{ip\Omega t}, \quad \mathbf{r}(t) = \mathbf{r}_{0,\perp} + \mathbf{v}_0 t + \delta\mathbf{r}_\perp(t) = \mathbf{r}_{0,\perp} + \mathbf{v}_0 t + \frac{c}{\Omega}\sum_{p\neq 0}^{\infty} \frac{\boldsymbol{\alpha}^{(p)}}{p^2} e^{ip\Omega t} \tag{8}$$

where $\boldsymbol{\alpha}^{(p)}$ is the dimensionless vector $\boldsymbol{\alpha}^{(p)} \equiv \frac{2e^2}{mc\gamma\Omega} \mathbf{u}_\perp^{(p)}$, the absolute value of which is the small parameter $|\alpha^{(p)}|<<1$, $\gamma$ is the Lorentz factor.

Substituting Eqs. (8) and (2) into Eq. (1), and multiplying it by $\mathbf{v}$, as scalar, we obtain the radiation power in the range $\omega_\lambda << \omega_{pe}$

$$P \equiv -\lim_{t\to\infty} \frac{1}{t}\int_0^t \mathbf{v}(t') \mathbf{F}(\mathbf{v}(t'),\mathbf{r}(t'),t') dt' =$$

$$= \frac{e^2 D}{2V_{cell}} \sum_{n=0}^{+\infty} \sum_{\lambda_j} \frac{1}{\left|v_0 - \frac{d\omega_\lambda}{dh}\right|_{\lambda=\lambda_j}} \left[ \left| \beta_0 g_{z,\lambda_j}^{(n)} + \frac{1}{2}\sum_{p\neq 0} \frac{\boldsymbol{\alpha}^{(p)}}{p}\left(\frac{D}{2\pi p}\nabla_\perp g_{z,\lambda_j}^{(n+p)} - i g_{\lambda_j}^{(n+p)}\right)\right|^2 + O\left(\left|\alpha^{(p)}\right|^2\right) \right] \tag{9}$$

Here $\omega_{\lambda_j}$ is the set of the frequencies found from the equations $\omega_\lambda - hv_0 = n\Omega$.

Eq.(8) shows that the radiation manifests itself in the interference between the Cherenkov-type radiation and the udulator-type radiation in the region $\omega_\lambda << \omega_{pe}$.

For the region $\omega_\lambda > \omega_{pe}$ it is interesting to consider the radiation emitted by a high-energy charged particle satisfying the condition $\omega_{pe} << 2\Omega\gamma^2$. Inserting Eqs.(8) and (3) into Eq.(1) and in analogy with Eq.(9) we obtain the power of the pure undulator-type radiation in the dipole approximation $k_\lambda \delta r_\perp(t) << 2\pi$

$$P_U \equiv -\lim_{t\to\infty}\frac{1}{t}\int_0^t \mathbf{v}(t')\mathbf{F}(\mathbf{v}(t'),\mathbf{r}(t'),t')dt' = \frac{4e^6}{3m^2c^3}\gamma^2 \sum_{p=1}^{p<<p_{\lim}} \left|\mathbf{u}_\perp^{(p)}\right|^2, \tag{10}$$

where the number of harmonics in the sum is limited by the condition of the smallness of the oscillation amplitude resulting in $p<<p_{\lim}=2\pi\gamma/\max\{\alpha^{(p)}\}$.

As follows from Eq.(10) the power grows as a square of the particle energy, so in the region $\omega >> \omega_{pe}$ the pure UR power can exceed the CR power emitted in the band $\omega << \omega_{pe}$.

**3.** It should also be stated that, if the bunch of $N$ electrons moves in the periodic structure and its longitudinal and transverse dimensions $\sigma_z$ and $\sigma_\perp$ satisfy the conditions $\sigma_z<<D/(2q\gamma^2)$ and $\sigma_\perp<<D/(2q\gamma)$, respectively, then the radiation with the frequencies $\omega < 2q\Omega\gamma^2$ would be coherent. Moreover, for the range $\omega_{pe} << \omega < 2q\Omega\gamma^2$ the UR power would be proportional to $N^4$.

$$P_U = \frac{4e^6 N^4}{3m^2c^3}\gamma^2 \sum_{p=1}^{q}\left|u_\perp(p)\right|^2 \tag{11}$$

Further, let us consider a case typical for rf linacs when the electron bunch dimensions $l_b$ satisfy the following relation

$$D/\gamma^2 << l_b < D \tag{12}$$

We suppose that the distribution of electrons is a random value in the small scale $\Delta l$ ($D/\gamma^2<\Delta l<<l_b$). So, for the ultrarelativistic beam satisfying the relation $\omega_{pe}<<2\Omega\gamma^2$ there is the spontaneous undulator-type radiation in the spectral range $\omega_{pe}<<\omega_\lambda$, the power of which can be written as

$$P_{Ub} = \frac{4e^6}{3m^2c^3}N^3 \sum_{p=1}^{p<<p_{\lim}} \int_{-\infty}^{\infty}d\tau \iint_{S_\perp} d^2r_\perp f_b(\boldsymbol{r}_\perp,\tau)\gamma(\boldsymbol{r}_\perp,\tau)^2\left|u_\perp^{(p)}(\boldsymbol{r}_\perp,\tau)\right|^2 , \tag{13}$$

where $f_b(\boldsymbol{r}_\perp,\tau)/v_0$ is the charge distribution density averaged over the small volume $\Delta l^3$, and normalized on the bunch charge $eN$, so $\int_{-\infty}^{\infty}d\tau \iint_{S_\perp} d^2r_\perp f_b(\boldsymbol{r}_\perp,\tau)=1$; $S_\perp$ is the cross-section of the flight space in the periodic structure; $\tau = t - z/v_0$; $\boldsymbol{u}_\perp^{(p)}(\boldsymbol{r}_\perp,\tau)$ is the $p$th spatial harmonic of the wake function determined as

$$\boldsymbol{u}_\perp^{(p)}(\boldsymbol{r}_\perp,\tau) \equiv -\frac{1}{e^2N}\int_0^\infty d\tau' \iint_{S_\perp} d^2r'_{0\perp} f_b(\boldsymbol{r}'_{0\perp},\tau-\tau')\boldsymbol{F}_\perp^{(p)}(\boldsymbol{r}_\perp,\boldsymbol{r}'_{0\perp},\tau'), \tag{14}$$

$\boldsymbol{F}_\perp^{(p)}(\boldsymbol{r}_\perp,\boldsymbol{r}_{0\perp},\tau)$ is the amplitude of the $p$th spatial harmonic of the transverse force acting on the charge $e$ having the transverse coordinate $\boldsymbol{r}_\perp$ and, at the distance $v_0\tau$, followed after the point charge $eN$ which has the transverse coordinate $\boldsymbol{r}_{0\perp}$.

**4.** For obtaining the quantitative values of the UR power let us consider a wakefied excitation by a monochromatic bunch of charged particles which moves in a weakly corrugated axially symmetrical waveguide with perfectly conducting walls. The periodic waveguide radius can be represented in the Fourier form

$$b(z) = b_0[1+\varepsilon(z)] = b_0[\ 1 + \sum_{p=-\infty}^{\infty}\varepsilon_p e^{i\frac{2\pi p}{D}z}\ ], \tag{15}$$

where $\varepsilon(z) <<1$ is the relative depth of the corrugation, $b_0$ is the average radius of the waveguide.

We solve the given task by the perturbation method which was expounded in [4]. The essential of this approach is to replace boundary conditions for the electric field $\boldsymbol{n}\times\boldsymbol{E}=0$, and magnetic field $\partial\boldsymbol{B}/\partial n=0$ ($\boldsymbol{n}$ is the normal to the waveguide surface) at $r = b(z)$ by the appropriate conditions on the values of $E_z$ and $\partial B_z/\partial r$ at $r = b_0$. These new conditions can be expressed in terms of the functions $b(z)$ and $db(z)/dz$ containing the small parameter $\varepsilon_p$. We will find the nonsynchronous harmonics of a traverse wake force which act on the bunch as wave pumping. At the start, following Ref. [4], we find the fields created by a point charge of $N$ electrons. Using the Maxwell's equations the transverse components of the fields are expressed in terms of the longitudinal components which satisfies the wave equations (see Eq.(1.18) and (1.19) in Ref. [4]). The wave equations are solved by the perturbation method representing the fields in the form of an expansion in terms of the powers of the small parameter $\varepsilon_p$.

The first order amplitude of the $p$th spatial harmonic of the transverse force acting on the charge $e$ having the transverse coordinates $r$, $\varphi$ and, at the distance $v_0\tau$, followed after the point charge $eN$ with the transverse coordinates $r=r_b$ and $\varphi=0$, is given by

$$F_r^{(p)}(r,\varphi,r_b,\tau) = -\frac{i8\pi e N p \varepsilon_p}{D}\sum_{m=0}^{\infty}\frac{(r_b)^m \cos(m\varphi)}{(1+\delta_{0,m})b_0^{m+1}}\sum_{s=1}^{\infty}\left\{A_{m,s}(r/b_0)e^{i\omega_{m,s,p}\tau} - B_{m,s}(r/b_0)e^{i\omega'_{m,s,p}\tau}\right\} \tag{16}$$

where $A_{m,s}(r/b_0) \equiv \dfrac{J'_m(\mu_{m,s} r/b_0)}{J'_m(\mu_{m,s})}$, $B_{m,s}(r/b_0) \equiv \dfrac{b_0}{r} \dfrac{m^2}{m^2 - \mu'^2_{m,s}} \dfrac{J_m(\mu'_{m,s} r/b_0)}{J_m(\mu'_{m,s})}$, in which

$\omega_{m,s,p} = \dfrac{\pi p c}{D}\left[1 + \left(\dfrac{D\mu_{m,s}}{2\pi p b_0}\right)^2\right]$ and $\omega'_{m,s,p} = \dfrac{\pi p c}{D}\left[1 + \left(\dfrac{D\mu'_{m,s}}{2\pi p b_0}\right)^2\right]$ are the frequencies of resonance modes excited by the point charge in the periodic structure; $\mu_{m,s}$ and $\mu'_{m,s}$ are the zeros of Bessel functions $J_m(\mu_{m,s}) = 0$ and $J'_m(\mu'_{m,s}) = 0$ respectively; $\delta_{0,m}$ is the Chronicler's symbol.

In the circumstances the longitudinal component of the synchronous harmonic of the electric field seen by the particle followed after the point charge is given by the expression

$$E_z^{(0)}(r,\varphi,\tau) = -\dfrac{4\pi e N}{cD} \sum_{m=0}^{\infty} \dfrac{(r_b r)^m \cos(m\varphi)}{(1+\delta_{0,m}) b_0^{2m}} \sum_{q=1}^{\infty} q |2\varepsilon_q|^2 \sum_{s=1}^{\infty} \left\{ \omega_{m,s,q} \cos(\omega_{m,s,q}\tau) - \dfrac{m^2 \omega'_{m,s,q} \cos(\omega'_{m,s,q}\tau)}{m^2 - \mu'^2_{m,s}} \right\} \quad (17)$$

A field within the bunch with an arbitrary charge distribution can be obtained by using Eq.(16) and Eq.(17) as Green's functions. To simplify the calculations we consider a filament bunch having a length $l_b$ and homogeneous charge distribution (averaged over the small volume $\Delta l^3$)

$$f_b(r,\varphi,\tau) = \dfrac{v_0}{2\pi r_b l_b}[H(\tau + l_b/2v_0) - H(\tau - l_b/2v_0)]\delta(r-r_b)\delta(\varphi), \quad (18)$$

where $H(\tau)$ is the Heaviside's step function, $H(\tau)=1$ for $\tau>0$ and $H(\tau)=0$ for $\tau<0$. Substituting Eqs.(18) and (16) into Eq.(14), and then substituting the found expression into Eq.(13), we obtain the total power of the spontaneous undulator-type radiation emitted by the bunch

$$P_{Ub} = \dfrac{4e^6}{3m^2c^3} \gamma^2 N^3 \left(\dfrac{4\pi}{b_0 D}\right)^2 \sum_{p=1}^{\infty} p^2 |2\varepsilon_p|^2 \sum_{m=0}^{\infty}\sum_{n=0}^{\infty} \dfrac{(r_b/b_0)^{m+n}}{(1+\delta_{0,m})(1+\delta_{0,n})} \times$$

$$\times \sum_{s=1}^{\infty}\sum_{q=1}^{\infty} \left\{ A_{m,s}\left(\dfrac{r_b}{b_0}\right) A_{n,q}\left(\dfrac{r_b}{b_0}\right) I(\omega_{m,s,p},\omega_{n,q,p}) + B_{m,s}\left(\dfrac{r_b}{b_0}\right) B_{n,q}\left(\dfrac{r_b}{b_0}\right) I(\omega'_{m,s,p},\omega'_{n,q,p}) - \right. \quad (19)$$

$$\left. - A_{m,s}\left(\dfrac{r_b}{b_0}\right) B_{n,q}\left(\dfrac{r_b}{b_0}\right) I(\omega_{m,s,p},\omega'_{n,q,p}) - B_{m,s}\left(\dfrac{r_b}{b_0}\right) A_{n,q}\left(\dfrac{r_b}{b_0}\right) I(\omega'_{m,s,p},\omega_{n,q,p}) \right\}$$

where $I(\omega_1,\omega_2) \equiv \dfrac{c^2}{\omega_1 \omega_2 l_b^2}\left[1 - \dfrac{\sin(\omega_1 l_b/c)}{\omega_1 l_b/c} - \dfrac{\sin(\omega_2 l_b/c)}{\omega_2 l_b/c} + \dfrac{\sin((\omega_1 - \omega_2)l_b/c)}{(\omega_1 - \omega_2)l_b/c}\right]$

Therewith the bunch power loss the associated with exciting the wakefield is given by

$$P_{wf} = -v_0 e N \int_{-\infty}^{\infty} d\tau \int_{0}^{b(z)} r dr \int_{0}^{2\pi} d\varphi f_b(r,\varphi,\tau) E_z^{(0)}(r,\varphi,\tau) = \dfrac{2\pi(eN)^2}{D} \sum_{p=1}^{\infty} p |2\varepsilon_p|^2 \sum_{m=0}^{\infty} \left(\dfrac{r_b}{b_0}\right)^{2m} \dfrac{1}{(1+\delta_{0,m})} \times$$

$$\times \sum_{s=1}^{\infty} \left\{ \omega_{m,s,p}\left(\dfrac{\sin(\omega_{m,s,p} l_b/2c)}{\omega_{m,s,p} l_b/2c}\right)^2 - \dfrac{m^2 \omega'_{m,s,p}}{m^2 - \mu'^2_{m,s}}\left(\dfrac{\sin(\omega'_{m,s,p} l_b/2c)}{\omega'_{m,s,p} l_b/2c}\right)^2 \right\} \quad (20)$$

**5.** Let us consider, as an example, the sinus-type corrugated waveguide with $\varepsilon(z) = 2\varepsilon_1 \sin(2\pi z/D)$ at $2\varepsilon_1 = 0.1$ and $b_0 = D$. To estimate the maximal magnitudes of the UR power we choose the distance of the bunch from the waveguide axis at the level of $r_b/b_0 = 0.9$. We will consider two variants. In the first variant an ultrarelativistic electron bunch, typical for SLAC [5] with $l_b = 500$ μm and $N = 4\times 10^{10}$, interacts with the periodic structure of millimeter dimensions $b_0 = D = 3$ mm. In the second variant all sizes of the structure and the bunch are reduced in ten times, so that: $b_0 = D = 0.3$ mm; $l_b = 50$ μm.

In Fig.1 represented are the dependences of the spontaneous UR power (lines 1 and 2) on the energy bunch and the similar dependences of the power loss associated with exciting the wakefield (lines 1' and 2'). The numbers 1, 1' correspond to the first variant and the numbers 2, 2' relate to the second variant. In the calculation we have taken into account the contribution of 120 resonance modes to the wakefield. As seen from Fig.1 there is the electron energy above of which the undulator-type radiation power exceeds the power loss associated with exciting the wakefield. The undulator-type radiation predominates at energies above 6.4 TeV for the first case (the millimeter structure) and higher than 80 GeV for the second variant (the submillimeter structures).

Fig.2 represents the dependences of the number of electrons in the bunch on their energy. In the region of $N$ and $mc^2\gamma$ parameters, these dependences determinate the limit above of which the spontaneous UR power exceeds the power of the wakefield excitation, $P_{Ub} > P_{wf}$. The lines 1 and 2 correspond to the variants 1 and 2 respectively.

**6.** In conclusion it should be noted that the undulator-type radiation power is proportional to $\gamma^2$. So, in the future high-energy electron rf linacs, in view of a natural beam deviation from the linac axis, because of the coherent betatron oscillation in a focussing system, the interaction of electrons with the spatial nonsynchronous harmonics of both an accelerating field [6] and a wakefield may result in the electron energy loss associated with the spontaneous undulator-type radiation.

Note, also, that the new radiation mechanism considered above can be used in undulators based on periodic structures without external fields, where the nonsynchronous wake-harmonics of an electron bunch act as wave pumping. These wakefield undulators require no magnetic fields or rf sources needed in present-day FEL. The development of such wakefield undulators with submillimeter periods may open the new possibilities for generating X rays without employing external periodic fields.

The author is grateful to Academician Ya.B.Fainberg for the proposed method of solution and for fruitful discussions and also to M.I.Ayzatsky, E.V. Bulyak, A.N.Dovbnia, V.F.Zhyglo and V.A.Kushnir for the series of valuable remark.

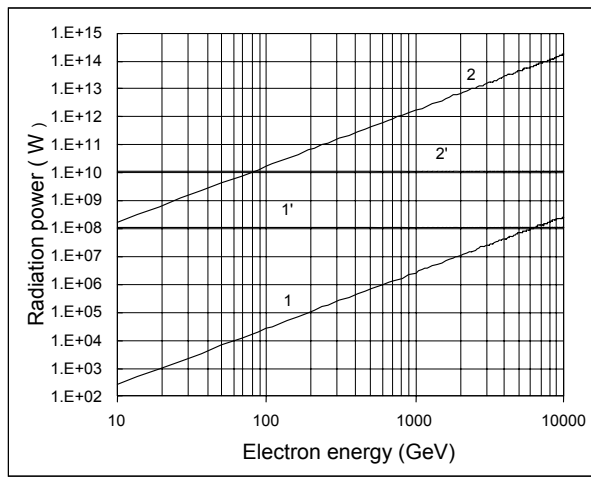
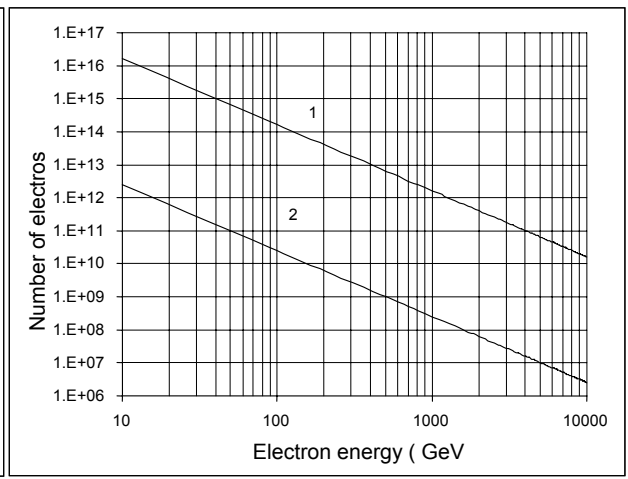

Fig.1. The radiation power v.s. the electron energy. 1, 2 – Undulator-type radiation; 1',2'- Cherenkov-type radiation.

Fig.2. The number of electros in the bunch v.s. the electron energy at the condition $P_{Ub} = P_{wf}$.